\documentclass[conference]{IEEEtran}
\ifCLASSINFOpdf
\else
\fi
\usepackage{graphicx,balance,epsfig,latexsym,bm,amsthm,url,amssymb,amsmath,qtree,caption,epstopdf,rotating,epsfig,subfigure,floatrow,algorithmic,footnote}
\usepackage[ruled,vlined]{algorithm2e}

\begin{document}
%
\title{NEDindex: A new metric for community structure in networks}

\author{\IEEEauthorblockN{Md. Khaledur Rahman }

\IEEEauthorblockA{Graduate Student at Dept. of CSE, BUET\\
Lecturer, Dept. of CSE, BRACU\\
Email: khaled@bracu.ac.bd}
}


%


\maketitle

\begin{abstract}
There are several metrics (\emph{Modularity}, \emph{Mutual Information}, \emph{Conductance}, etc.) to evaluate the strength of graph clustering in large graphs. These metrics have great significance to measure the effectiveness and they are often used to find the strongly connected clusters with respect to the whole graph. In this paper, we propose a new metric to evaluate the strength of graph clustering and also study its applications. We show that our proposed metric has great consistency which is similar to other metrics and easy to calculate. Our proposed metric also shows consistency where other metrics fail in some special cases. We demonstrate that our metric has reasonable strength while extracting strongly connected communities in both simulated (\emph{in silico}) data and real data networks. We also show some comparative results of our proposed metric with other popular metric(s) for Online Social Networks (OSN) and Gene Regulatory Networks (GRN).
\end{abstract}


%
\IEEEpeerreviewmaketitle

\section{Introduction}
\label{introduction}
Analysis of graph is very important in OSN, GRN, Distributed Computing Systems (DCS), Multi-Core Architecture (MCA) and so on. In OSN, we can easily represent the interactions among users by a graph where we can map users as nodes and interactions among users as edges. Similarly, we can reproduce a graph from other fields (GRN, DCS, MCA) as well. It is important to note that all users do not communicate with all other users in OSN. They form community and interact with other users in that community. All the \emph{regulators} do not interact with all other \emph{genes} in GRN. A set of \emph{regulators} interacts with a particular set of \emph{genes} i.e., they also form groups or communities. Thus, effective graph clustering has become an important research area where several groups or communities are formed in large graph. A large body of research works have been carried out in these areas \cite{may2001,redner1998,newman2001,jeong2000}. Notably, some research works have been done to measure the effectiveness of such graph clustering or graph partitioning \cite{albert2002,newman2004}.

\begin{figure}[h]
\centering
\caption{A graph with two clusters.}\label{fig1}
\includegraphics[width=0.9\textwidth]{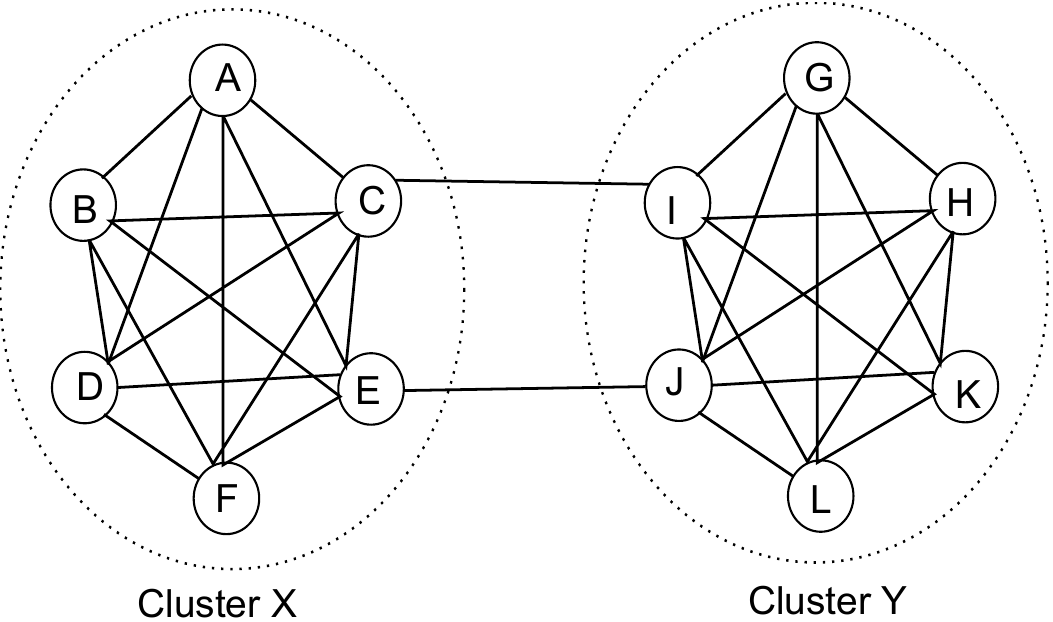}
\end{figure}

An undirected graph, $G$ is represented as $G = (V, E)$, where $V$ is the total number of vertices (nodes) and $E$ is the total number of edges (link between any two nodes) in graph $G$. Clusters are represented as $C = \{C_1, C_2, \ldots, C_n\}$, where $C_i$ is a cluster that consists of a set of nodes and associated edges from the graph, $1\le i \le n$. In Figure \ref{fig1}, we see a graph, where $V = \{ A, B, C, D, E, F, G, H, I, J, K, L\}$ and $E=\{ AB, AC, AD, \ldots$ \emph{all links between any two vertices}$\}$. There are two clusters, $C= \{C_1, C_2\}$ such that $C_1=\{A, B, C, D, E, F\}$ and $C_2=\{G, H, I, J, K, L\}$. We have shown $C_1$ and $C_2$ by dotted circles in Figure \ref{fig1} as `Cluster X' and `Cluster Y'. We use the term \emph{component} to represent all the vertices and associated edges of a cluster.

Any size of clusters provide necessary information and these are not helpful in general to analyze complex networks. If we want to break a given network into clusters, we do not know how many clusters there are going to be and it is true that all clusters should not be the same size. If we consider Facebook, there are billions of users (nodes) as well as billions of interactions (edges) and we do not know how many clusters or communities are there. While clustering, we do not even need to minimize inter-community edges as more such edges may be present between large communities compared to small ones. Sometimes, a small strongly connected community is superior to other weakly connected communities. So, effective clustering is a difficult but important task in large graphs.


Now, the question is how can we cluster such large graphs of OSN or GRN and measure the effectiveness of such clustering. For example, what would be the problem if we clustered the graph in Figure \ref{fig1} as $C = \{C_1, C_2, C_3\}$ such that $C_1=\{A, B, C, D\}$, $C_2=\{E, F, J, L\}$ and $C_3=\{G, H, I, K\}$? It seems that there would have weekly connected cluster(s). We observe in Figure \ref{fig1} that two clusters have greater strength than three clusters. So, we need some metrics that evaluate such strength. A lot of research works have been done to discover such metrics. In this paper, we also focus on presenting a metric that is more effective and efficient to calculate. More specifically, we propose a graph clustering metric named Node-Edge-Degree index (NEDindex) and successfully apply this on both simulated data and real-world data. We use popular graph theories to estimate the value of NEDindex which varies between $[0-1]$. We also show that NEDindex has greater consistency than some other metrics. In fact, NEDindex gives consistent measurement where some popular metrics fail to evaluate effectiveness. In addition, we apply NEDindex in some real-world application like community detection in social networks and sub-network classification in gene regulatory networks. In the rest of the paper, we use cluster, community, sub-network and group interchangeably.

We organize rest of the discussion as follows. We give some existing metrics in section \ref{literature}. In section \ref{solution}, we discuss our proposed metric and show how to calculate it. We give experimental results in section \ref{experiment} and provide two real world applications in section \ref{app}. Finally, we conclude briefly in section \ref{conclusion}.

\section{Related Works} 
\label{literature}
In the literature, there are lots of works regarding graph clustering. As we are more concerned with determining strength of a cluster in a graph, a more useful technique is used in social network analysis known as hierarchical clustering. In this technique, clusters are discovered from natural divisions depending on some similarity metrics or strong interactions between nodes. We can divide these into two broad classes namely agglomerative and divisive \cite{scott2000}. This classification depends on whether we add or remove edge(s) to or from the social graphs. Sometimes, we may observe some built-in similarity. For example, in filmography, network of collaborations between film celebrities has been studied in \cite{watts1998,amaral2000} where actress/actors are interlinked if they perform in the same film. One can also measure the similarity by how many movies in which actress/actors act together \cite{marchiori2000}.


In \cite{newman2004}, authors focus on divisive methods. This method actually starts with the network of interest and try to locate the weakly connected pairs of nodes and then unlink the edges between them. They use \emph{betweenness} measure of graph theory. The simplest example of such measure is that based on shortest (geodesic) paths: they find the shortest paths between all pairs of vertices and count how many run along each edge. They call it \emph{modularity} and according to authors, this measure was first introduced by Anthonisse in a never-published technical report in 1971 \cite{anthonisse1971}. Anthonisse named it ``rush", but they use the term edge betweenness, since the measurement is a natural generalization to edges of the strongly connected (vertices) betweenness measure of Freeman \cite{freeman1977}. Now, \emph{modularity} has become a very popular metric in graph clustering. In \cite{danon2005}, authors provide another reliable metric named \emph{Normalized Mutual Information} (NMI) which is a measure of similarity for groupings taken from information theory.

In this work, we focus more on \emph{degree centrality} of a graph. This technique has been relatively little explored in the previous literature, either in OSN or GRN, but, as we will see, seems to provide a lot of consistency.

\section{The NEDindex metric}
\label{solution}
In this section, we present all the necessary equations to estimate NEDindex metric. For the sake of simlicity, here, we use undirected unweighted graph to calculate NEDindex. But, we can also calculate NEDindex for other kinds of graph like directed-weighted, undirected-weighted and directed-unweighted. At first, we give a popular formula from graph theory in Equation \ref{eqn1}. It says that the total degree of an undirected graph is equal to twice the number of edges in that graph.
\begin{equation}
\label{eqn1}
D(G) = 2\sum_{1\leq i \leq n} e_i
\end{equation}
Here, $e_i$ represents edge(s) in graph $G=(V, E)$ such that $E=\{e_1, e_2, \ldots, e_n\}$, $1\leq i \leq n$. We present Equation \ref{eqn2} to calculate $NED(C)$, where $C$ is a cluster in graph $G$. 
\begin{equation}
\label{eqn2}
NED(C) = \frac{|V_c| + |E_c| + D(C) }{|V_c|+{|V_c|\choose 2} + D(G, V_c)}
\end{equation}
Here, $|V_c|$ is the number of vertices in cluster $C$, $|E_c|$ is the number of edges among all vertices ($V_c$) in cluster $C$, $D(C)$ is the total degree of cluster $C$ considering $E_c$ edges and $D(G, V_c)$ is the total number of degree of $V_c$ vertices in original graph, $G$. Note that ${|V_c|\choose 2}$ is the total number of edges in a complete graph of $V_c$ vertices such that $V_c\subseteq V$. Finally, we calculate NEDindex using Equation \ref{eqn3} as follows.
\begin{equation}
\label{eqn3}
NEDindex = \sum_{1\leq i \leq n}\frac{NED(C_i)*D(C_i)}{D(G)}
\end{equation}
Here, $C_i$ represents clusters where, $n$ is the total number of clusters in graph $G$. 

\begin{figure}[H]
\centering
\caption{A simple graph with three clusters.}\label{fig2}
\includegraphics[width=0.9\textwidth]{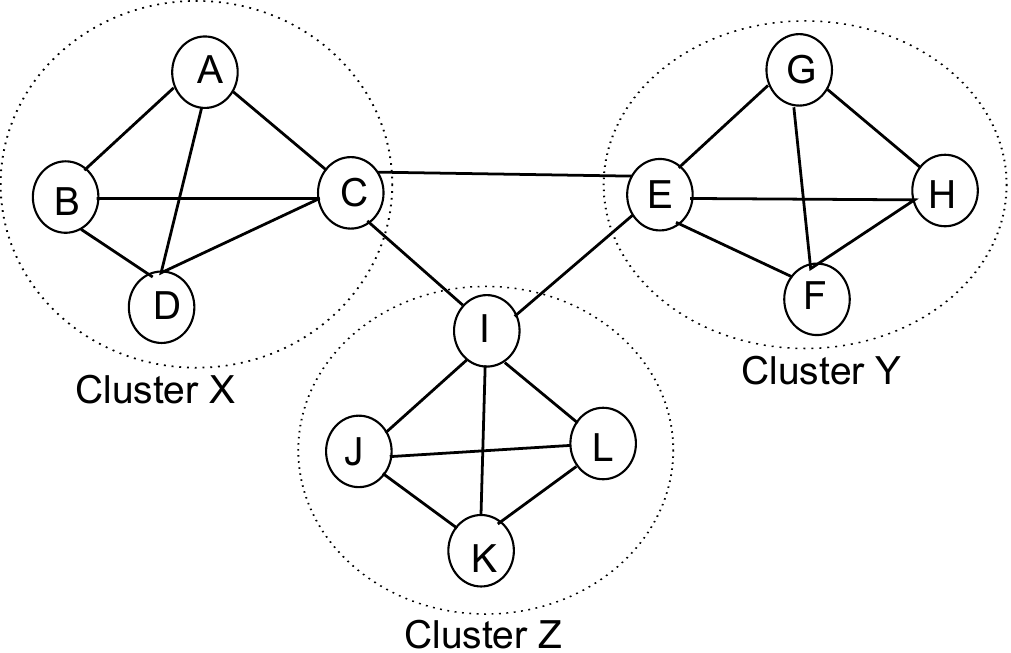}
\end{figure}

We show a graph in Figure \ref{fig2} as an example of three clusters. In this Figure, we denote `Cluster X', `Cluster Y' and `Cluster Z' as $C_1$, $C_2$ and $C_3$ respectively. So, we calculate NEDindex for this graph (Figure \ref{fig2}) as follows:
\begin{equation*}
\label{eqn4}
NED(C_1) = \frac{4+6+12}{4+6+14}=\frac{22}{24}=0.9167 
\end{equation*}
Similarly, we get $NED(C_2)=NED(C_3)=0.9167$.
\begin{equation*}
\label{eqn5}
NEDindex = \frac{0.9167*(12+ 12 + 12)}{42}=\frac{33.0012}{42}=0.7857 
\end{equation*}

The value of NEDindex will vary between 0 to 1. When the value of NEDindex is more close to 1, then all the nodes of that clusters are more strongly connected with respective to the whole graph. Similarly, nodes of clusters are weekly connected if value of NEDindex is close to 0. According to \cite{newman2004}, \emph{modularity} metric is calculated using following formula, where $Q$ is modularity.

\begin{equation*}
\label{eqn6}
Q = \frac{1}{2*E}\sum_{i,j}[A_{ij}-\frac{d_id_j}{2*E}]\delta(s_i,s_j)
\end{equation*}
Here, $E$ indicates the number of edges with respect to adjacency matrix $A$, $d_i$ indicates the degree of node $v_i$, $s_i$ indicates the community membership of node $v_i$ and $d(s_i, s_j) = 1$ if $s_i = s_j$.

The value of \emph{modularity} metric generally varies between [-1,+1] but sometimes it also varies between [0, 1] based on the representation. The value of NMI also varies between [0-1]. In case of \emph{modularity} or \emph{NMI}, the same information holds true about performance i.e., higher value indicates strongly connected community and lower value indicates weakly connected community. For the graph shown in Figure \ref{fig2}, we get the value of \emph{modularity} metric as  0.5238.

\begin{figure}[H]
\centering
\caption{Consistency of NEDindex with \emph{modularity} metric for graph shown in Figure \ref{fig2}.}\label{comp1}
\includegraphics[width=0.9\textwidth]{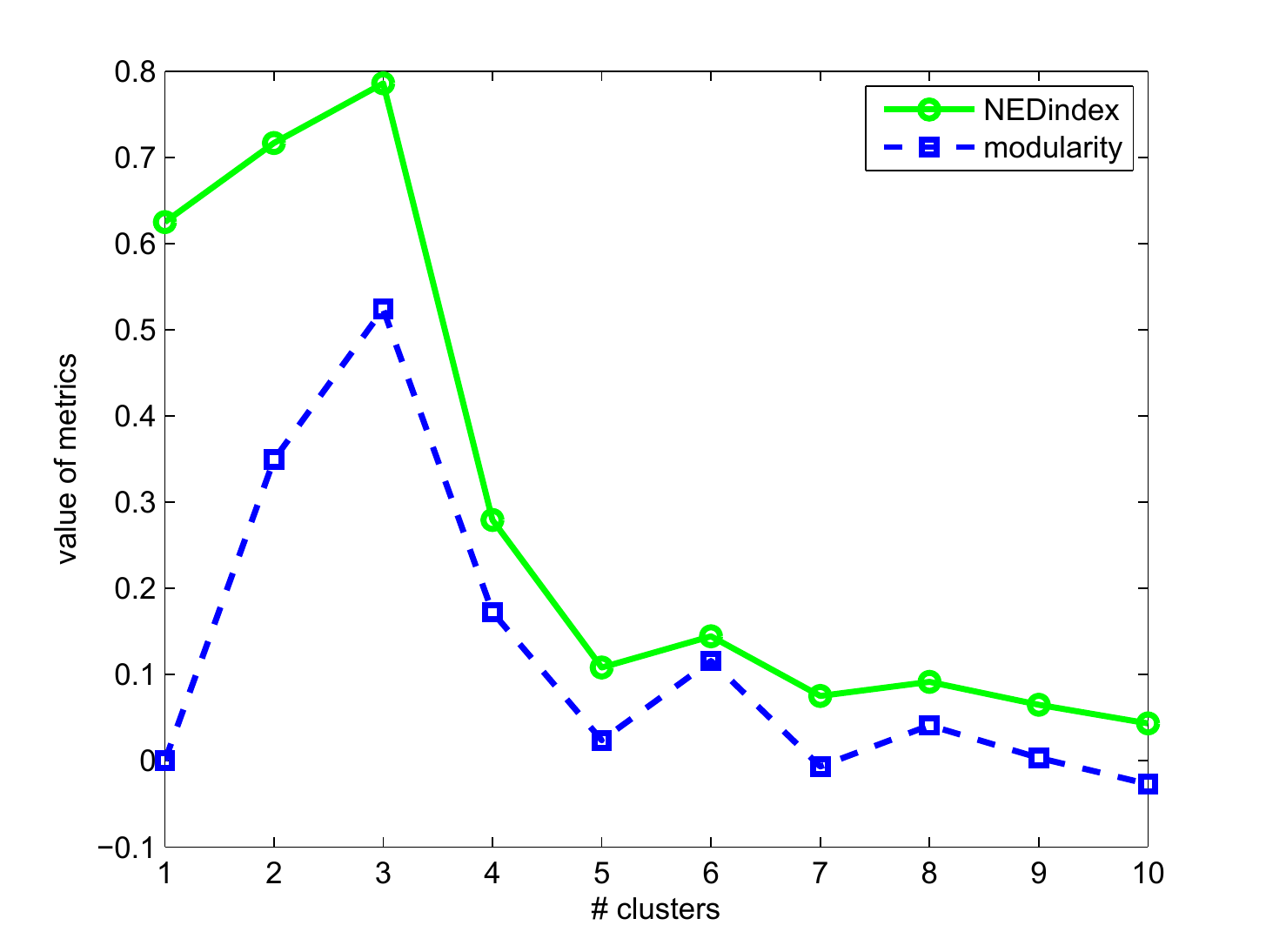}
\end{figure}

\begin{figure}[H]
\centering
\caption{Variation of NEDindex with \emph{modularity} metric for different combination of three clusters from graph shown in Figure \ref{fig2}. In each combination, a cluster is formed taking different number of nodes.}\label{comp2}
\includegraphics[width=0.9\textwidth]{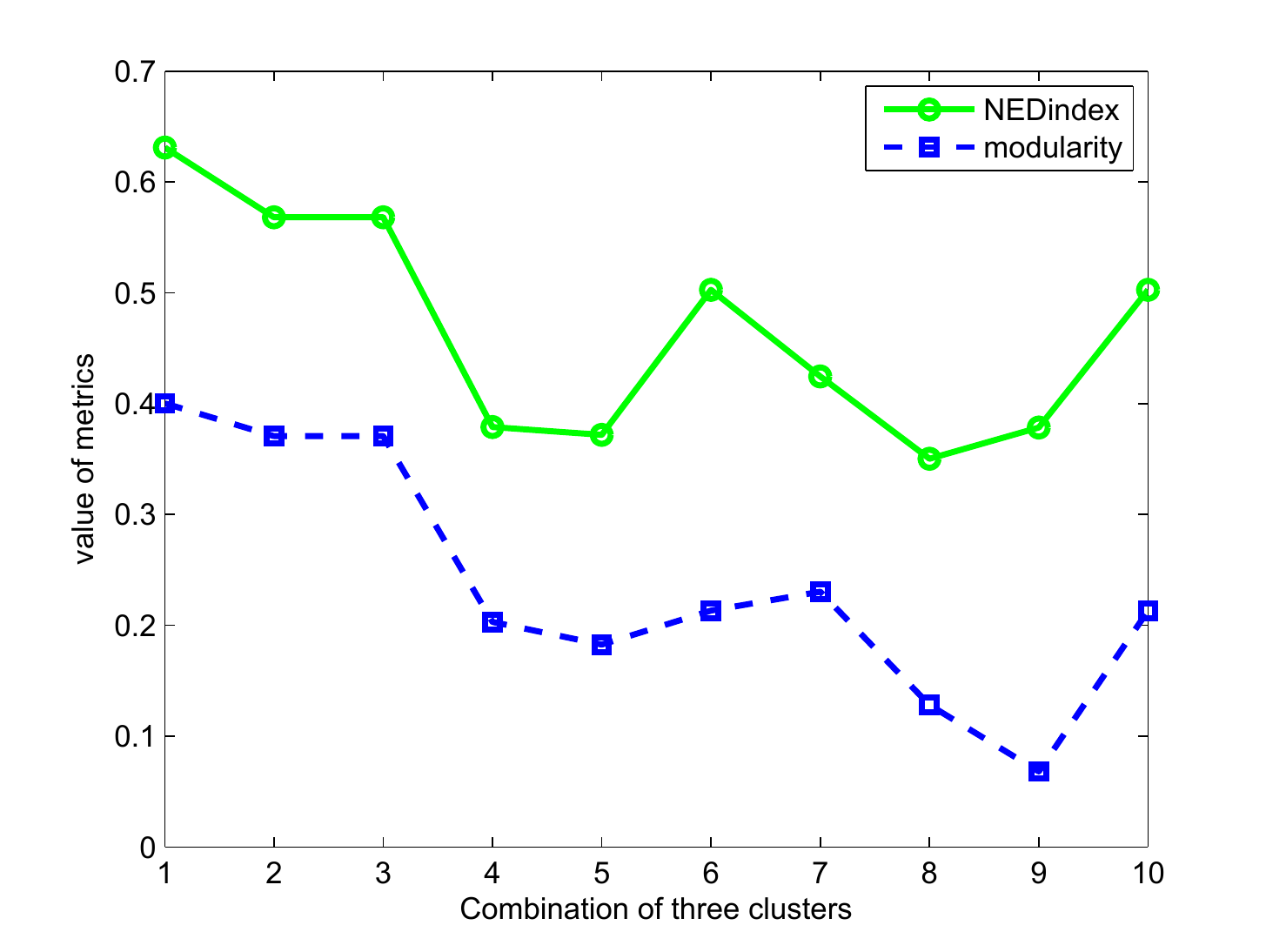}
\end{figure}

\section{Experimental Results}
\label{experiment}
\begin{figure}[H]
\centering
\caption{An example graph with three clusters.}\label{fig3}
\includegraphics[width=0.9\textwidth]{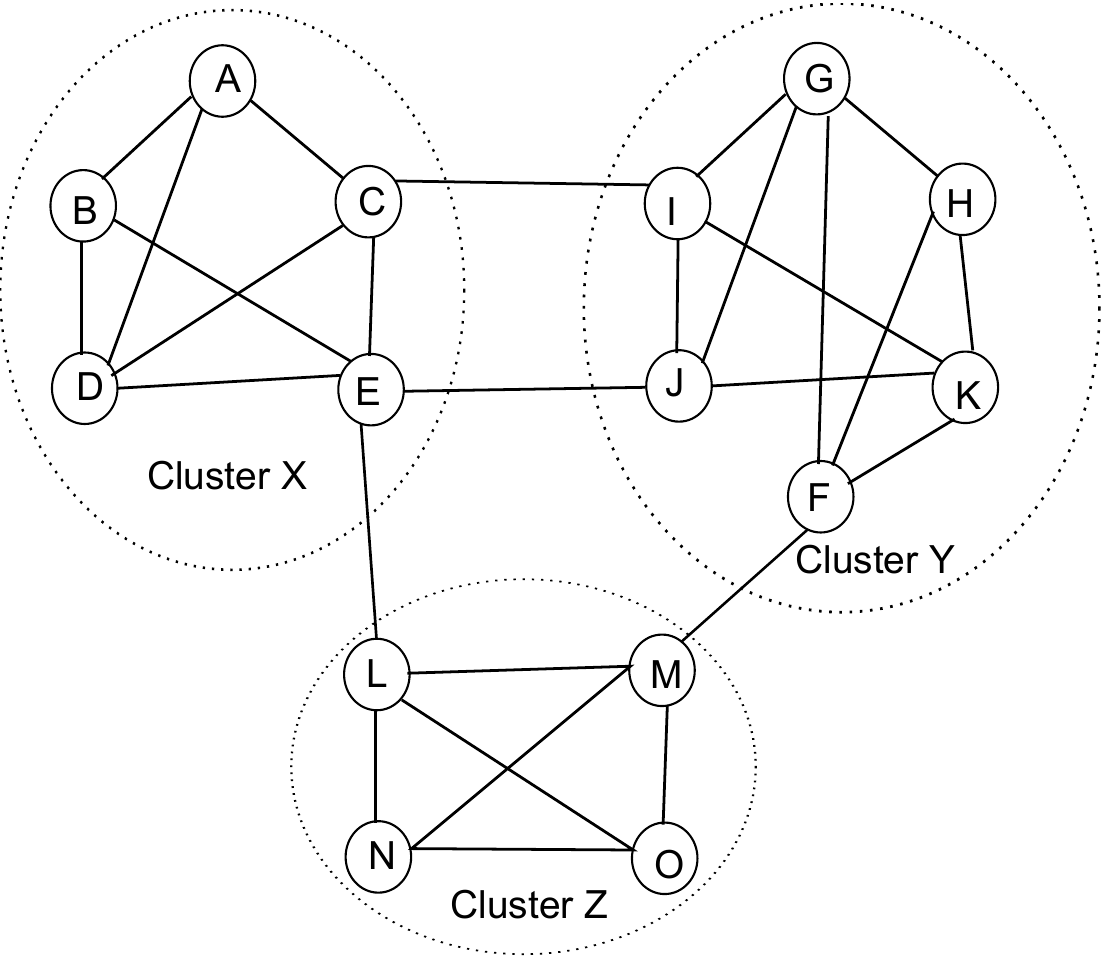}
\end{figure}

In this section, we illustrate experimental results on our proposed metric. To conduct experiments, we implement NEDindex metric in Matlab language. We conduct all the experiments in a laptop PC configured as 4GB RAM, Intel Corei3 @ 1.9GHz  and 64-bit Windows 8. First, we calculate NEDindex value for two simple graphs shown in Figures \ref{fig2} and \ref{fig3}. Then, we estimate the value of NEDindex for Karate club network given from \cite{zachary1977}. In both cases, we show corresponding value of \emph{modularity} metric. Finally, we present two exceptional results where \emph{modularity} metric does not show consistency but NEDindex performs well.

In Figure \ref{comp1}, we show the measurement of both NEDindex and \emph{modularity} for various size of clusters based on Figure \ref{fig2}. We observe that, we get peak value in both cases for the cluster of size 3. To investigate further, we keep the cluster size as 3 but take different number of nodes in different clusters. We see that the value of NEDindex goes up and down with strong clusters and weak clusters respectively. We report the results in Figure \ref{comp2}.

\begin{figure}[H]
\centering
\caption{Consistency of NEDindex with \emph{modularity} metric for graph shown in Figure \ref{fig3}.}\label{comp3}
\includegraphics[width=0.9\textwidth]{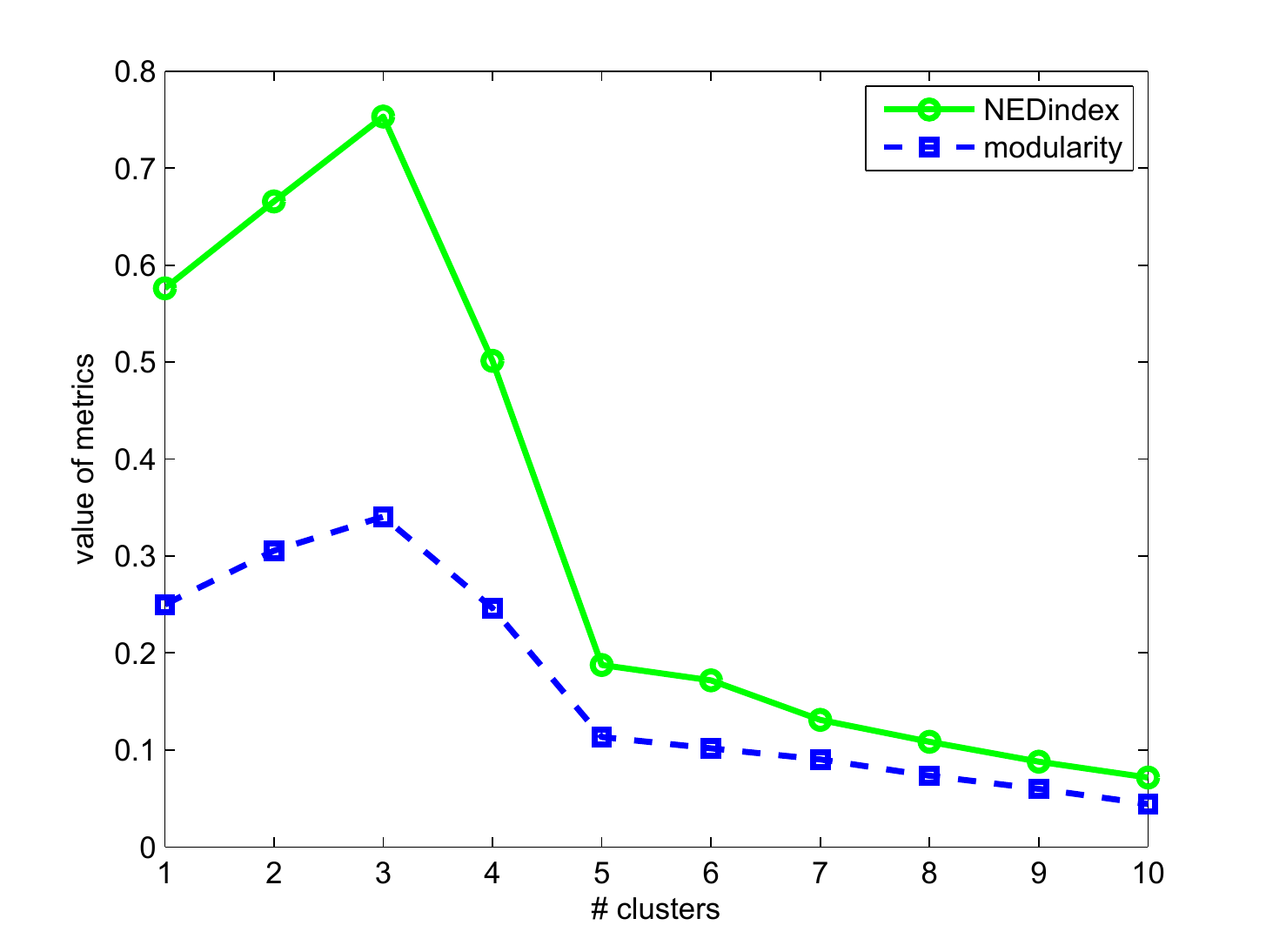}
\end{figure}
We conduct experiments on another graph shown in Figure \ref{fig3}. From this graph, we can easily observe that there are three strongly connected clusters, where $C_1=\{ A, B, C, D, E\}$, $C_2=\{F, G, H, I, J, K\}$ and $C_3=\{L, M, N, O\}$. However, while doing experiments, we vary the size of clusters. We report experimental results in Figure \ref{comp3}. We clearly see that experimental results support our observation i.e., $C=\{C_1, C_2, C_3\}$ forms strongly connected clusters and we get the peak value for that $C$.

\begin{algorithm}
\caption{Metrics Comparison}
\label{algo1}
\begin{algorithmic}
\REQUIRE \textbf{G(V, E)}
\STATE G(V, E) is an adjacent matrix of $V$ vertices and $E$ edges.
\STATE $L\rightarrow$ \textbf{\textit{linkage}}$(G)$
\FOR{$i \rightarrow $ 1 to $|V|$}
\STATE  $CM \rightarrow$ \textbf{\textit{cluster}}$(L, `maxclust', i)$
\STATE $C\rightarrow \emptyset$
	\FOR{$j \rightarrow $ 1 to \textbf{\textit{length}}$(CM)$}
		\STATE $C(CM(j)) = C(CM(j)) \cup \{j\}$
	\ENDFOR
	\STATE Calculate NEDindex using $C$ and $G$
	\STATE Calculate \emph{modularity} using $C$ and $G$
	\STATE Calculate \emph{NMI} using $C$ and $G$
\ENDFOR

\end{algorithmic}
\end{algorithm}

\begin{figure}[h]
\centering
\caption{Consistency of metrics for Karate network of \cite{zachary1977}.}\label{comp4}
\includegraphics[width=0.9\textwidth]{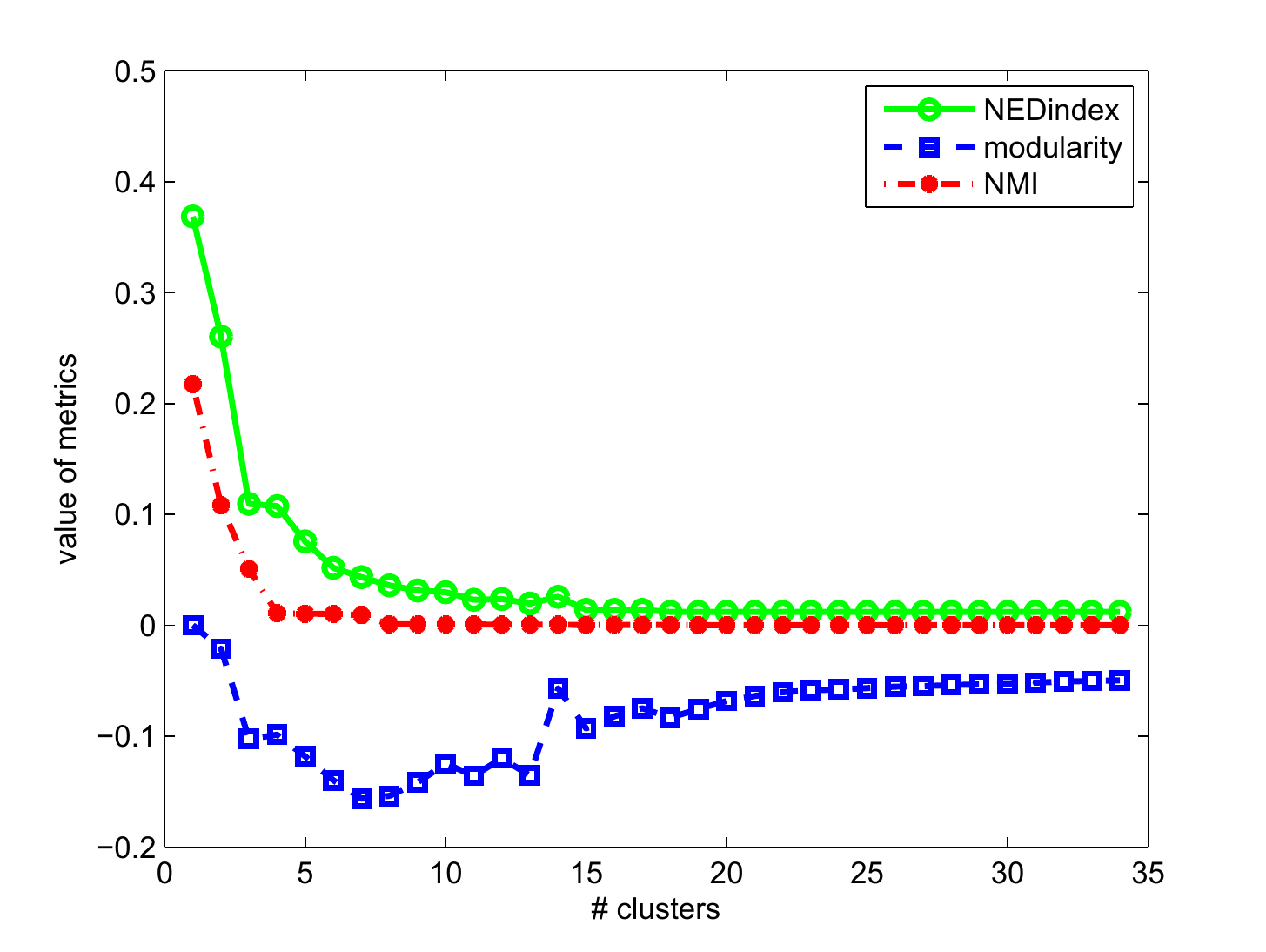}
\end{figure}

Now, we consider Karate club network from \cite{zachary1977}. There are 34 vertices and 78 edges. We conduct random experiment on Karate network varying cluster size (also called number of clusters in a graph) from 1 to 34. For each of the cluster size, we use \textit{\textbf{linkage}} and \textit{\textbf{cluster}} functions of Matlab language and generate components for each of the clusters. We use Algorithm \ref{algo1} to conduct experiments and compare different metrics. Input of the algorithm is an adjacent matrix. According to our computation, we consider that the maximum number of clusters will be the total number of vertices in the graph. We calculate all three metrics (NEDindex, \emph{modularity}, \emph{NMI}) for the same size and components of clusters. We run the experiments five times for each size of cluster and take average value of each metric. We report the results in Figure \ref{comp4}. We see that NEDindex is consistent with increasing number of clusters. From Figure \ref{comp4}, we also get that clusters become weakly connected when number of clusters increase unnecessarily.

\begin{figure}[h]
\centering
\caption{Two exceptional kinds of graphs (Wheel, Complete) where inconsistency in measurement may arise}\label{fig4}
\includegraphics[width=0.9\textwidth]{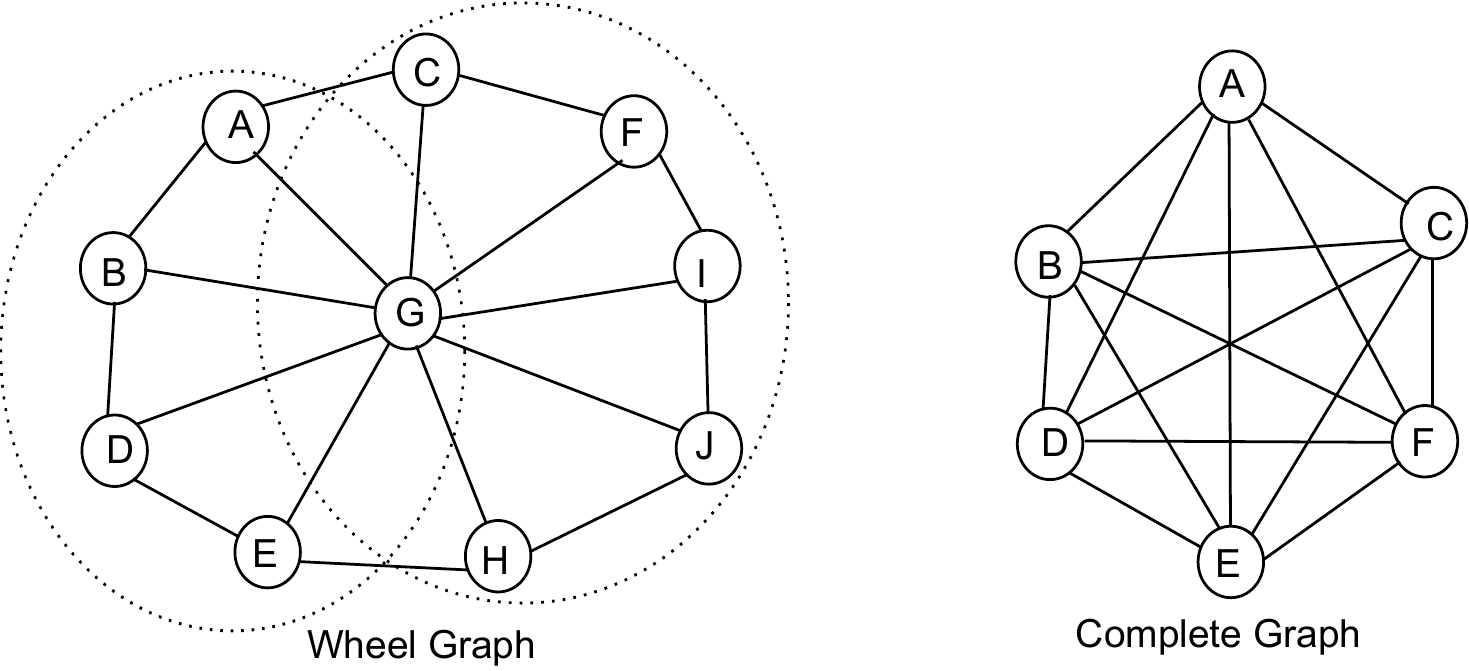}
\end{figure}

\begin{figure}[h]
\centering
\caption{Comparison of NEDindex and \emph{modularity} for wheel graph of Figure \ref{fig4}.}\label{figres1}
\includegraphics[width=0.9\textwidth]{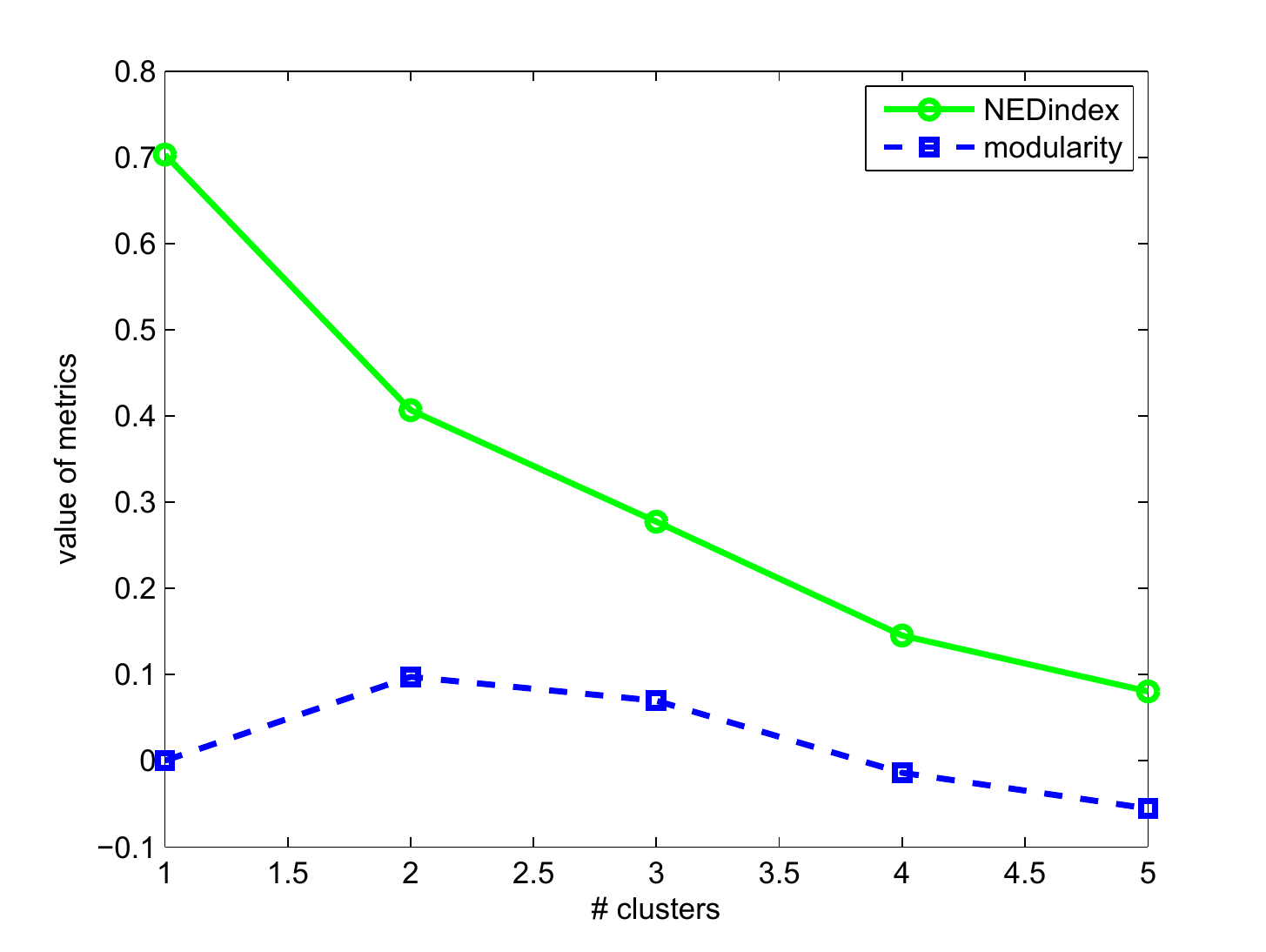}
\end{figure}

\begin{figure}[h]
\centering
\caption{Comparison of NEDindex and \emph{modularity} for complete graph of Figure \ref{fig4}.}\label{figres11}
\includegraphics[width=0.9\textwidth]{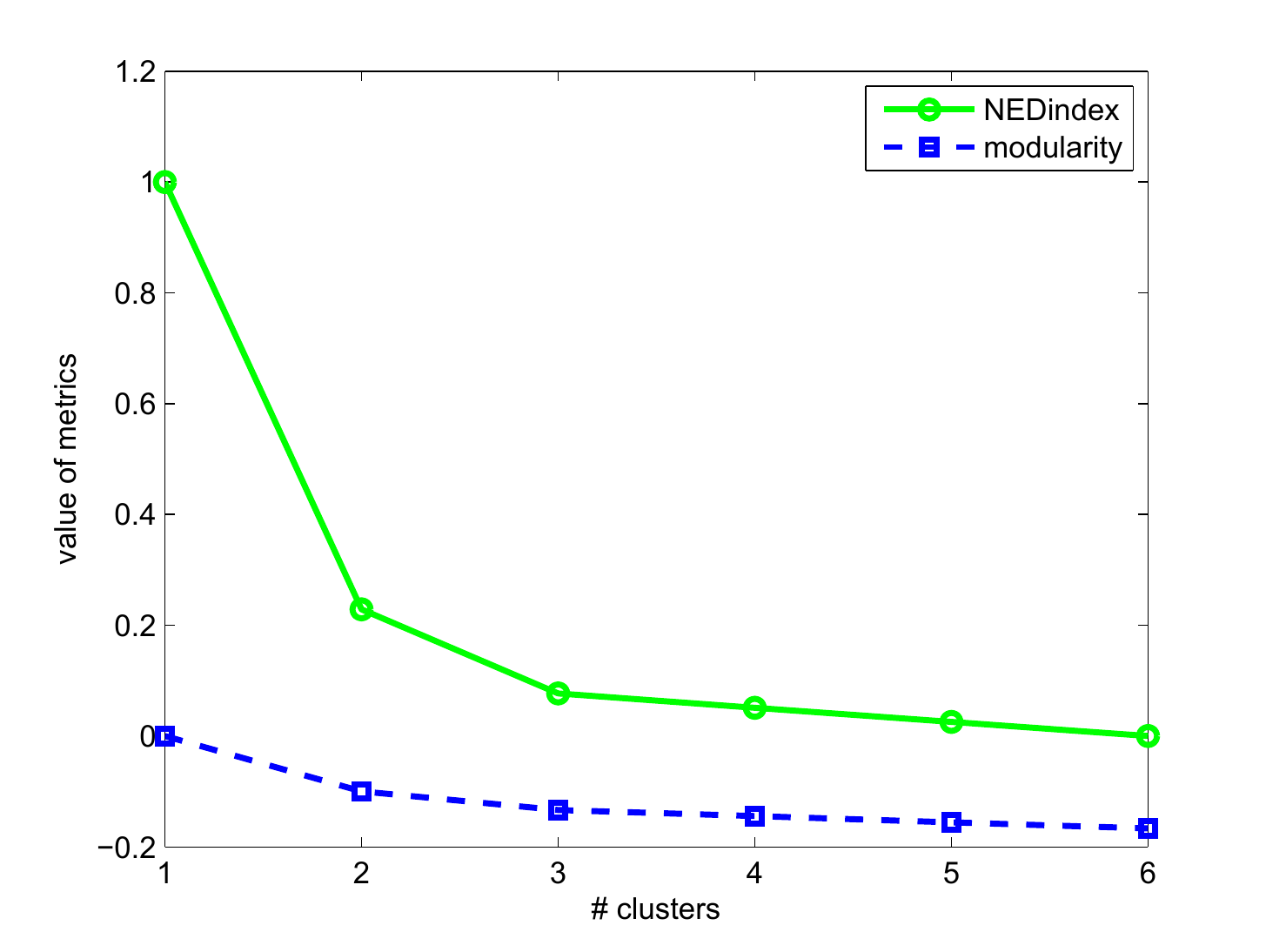}
\end{figure}

In Figure \ref{fig4}, we also show two exceptional graphs where \emph{modularity} seems inconsistent. For the complete graph shown in Figure \ref{fig4} (right), when we consider the cluster size 1, the \emph{modularity} metric is very close to 0. But, it should be close to 1 as cluster is strongly connected. After that the value of \emph{modularity} metric decreases with increasing number of cluster size. Our NEDindex is very close to 1 if the size of cluster is 1 for a complete graph. This shows high consistency while evaluating strongly connected components. After that the value of NEDindex decreases with increasing number of cluster size which also shows consistency. For the wheel graph shown in Figure \ref{fig4} (left), we also see the same characteristics for NEDindex. In case of \emph{modularity} metric, we see that the value of measurement increases when cluster size is 2. But, that should not be the usual case because the strength of connectedness is higher for cluster size 1 rather than 2. From cluster size 3 to upwards, the value of \emph{modularity} metric decreases again. We report the results of these two special cases in Figure \ref{figres1} and \ref{figres11}.

\subsection{Applications}
\label{app}

\begin{figure}[h]
\centering
\caption{Comparison of NEDindex, \emph{modularity} and \emph{conductance} for a benchmark graph in \cite{snap}. We use here \textbf{\textit{k-means}} clustering technique. }\label{figres2}
\includegraphics[width=0.9\textwidth]{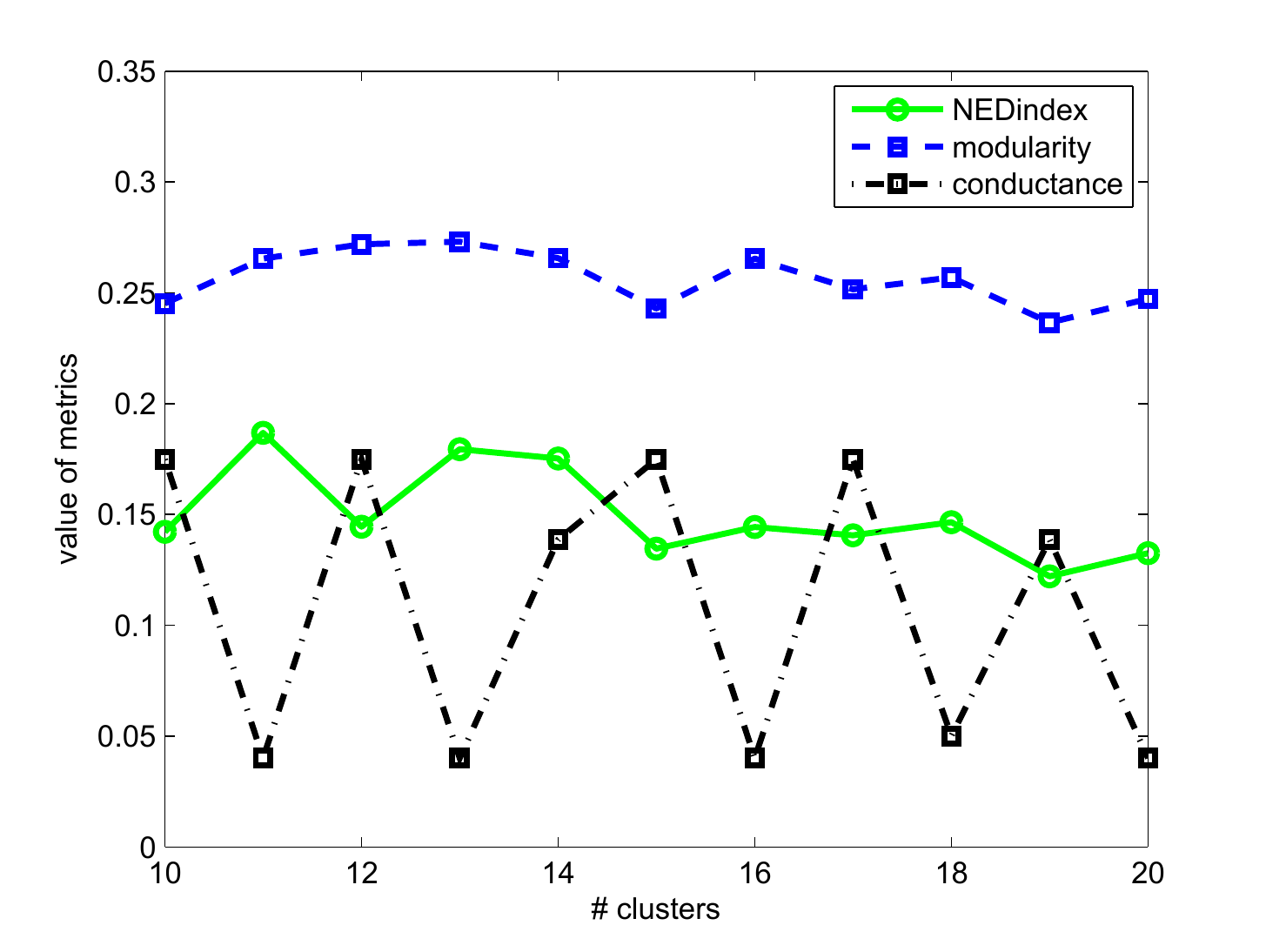}
\end{figure}

\subsubsection{Community Detection in Social Networks}
\label{commdetect}
\begin{figure}[h]
	\centering
    \caption{Comparison of NEDindex, \emph{modularity} and \emph{conductance} for a benchmark graph in \cite{snap}. We use here Algorithm \ref{algo1}.}
	\includegraphics[width=0.9\textwidth]{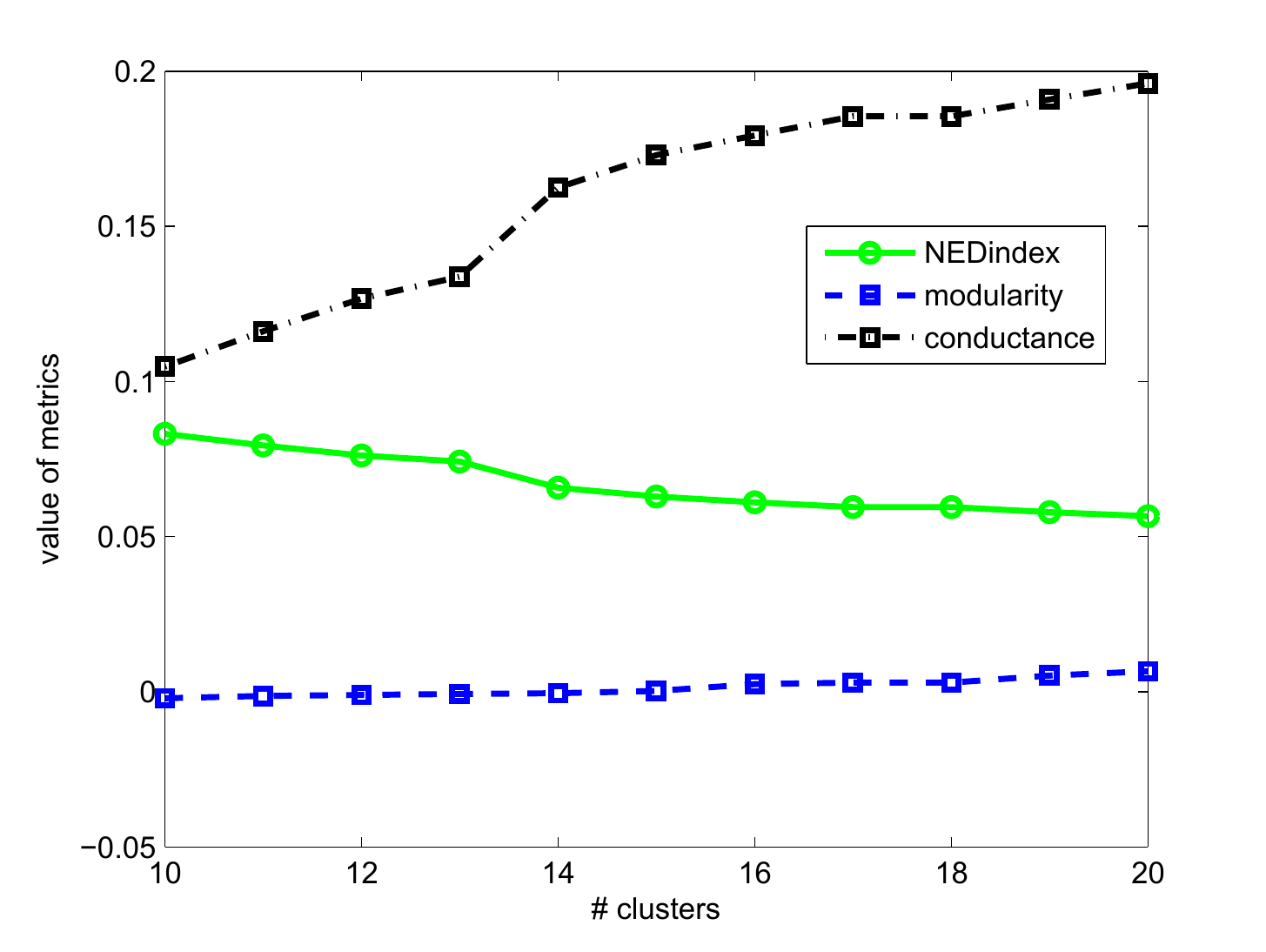}
    \label{figres3}
\end{figure}

Detecting strongly connected communities in social networks has become an area of interests in graph clustering. A large bodies of research have been carried out in this area. Sometimes, community detection algorithm consider overlapping clusters and sometimes it considers non-overlapping clusters. In \cite{linkscan}, authors propose an overlapping community detection technique. In \cite{empirical}, authors show comparative analyses of different methods of community detection algorithms. They also provide extensive discussions on performance metrics (\emph{conductance, exxpansion, internal density, cut ratio, modularity ratio,} etc). There are several benchmark datasets on social networks specially for large networks in \cite{snap}. So, we can easily apply NEDindex metric to any of them to detect strongly connected clusters in large graphs of social networks. We select facebook network (undirected graph) and use NEDindex metric to evaluate the strength of communities. There are 347 nodes and 5038 edges in the selected graph. We vary cluster size from 10 to 20. We measure other metrics as well and report comparative results in Figure \ref{figres2}. For Figure \ref{figres2}, we use \textbf{\textit{k-means}} clustering algorithm of Matlab language and for Figure \ref{figres3}, we use Algorithm \ref{algo1}. Here (in Figure \ref{figres2}), we see that \emph{conductance} fluctuates with the changes of cluster size. We also see that our proposed NEDindex shows consistency.

\subsubsection{Discovering Sub-network in Gene Regulatory Network}
\label{grn}
\begin{figure}[h]
\centering
\caption{Comparison of NEDindex with \emph{modularity} for GRN network of \cite{dream}.}\label{compl}
\includegraphics[width=0.9\textwidth]{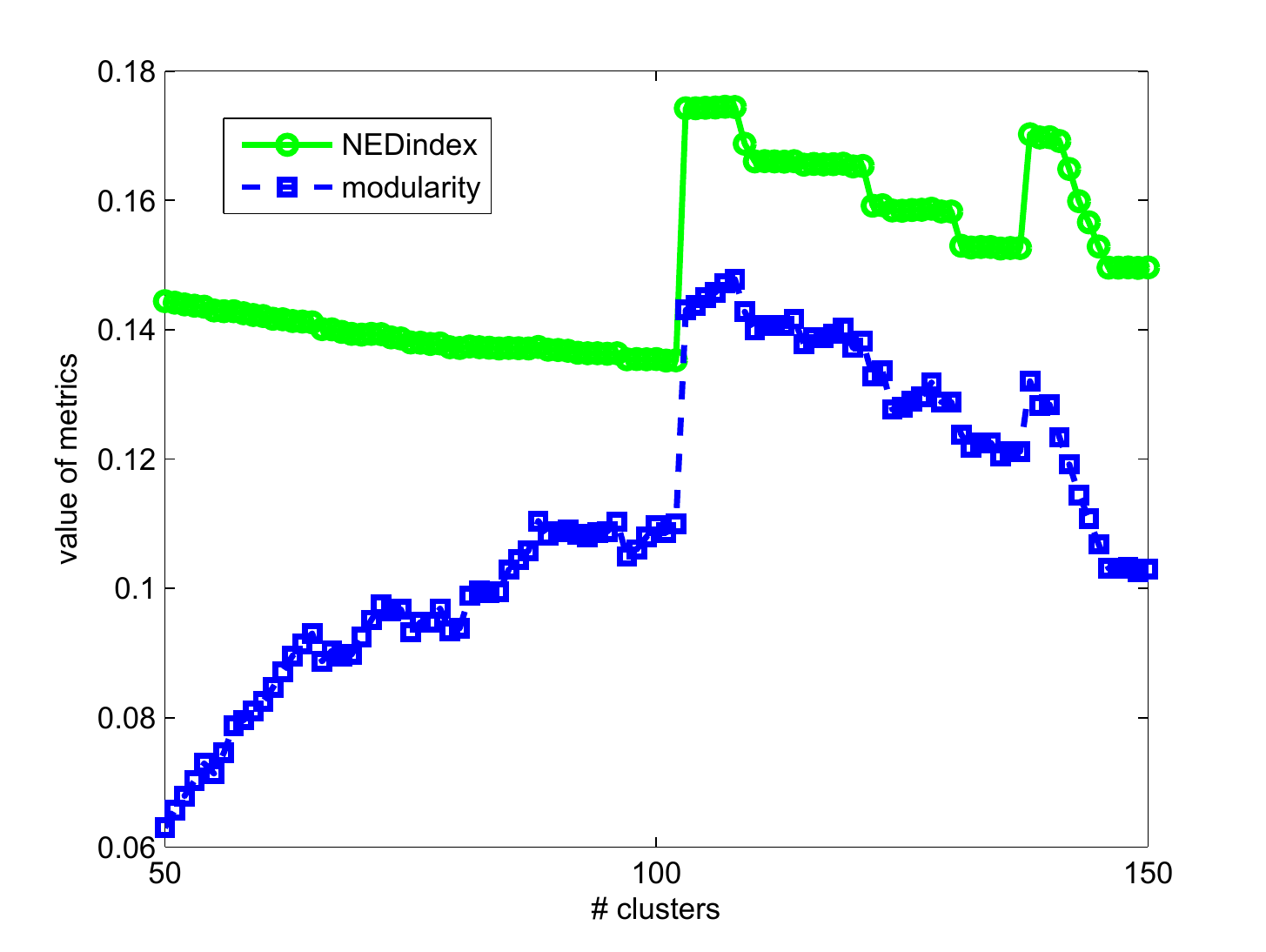}
\end{figure}

Re-construction of GRN is a highly growing field of applied bioinformatics. There are lots of works in this field where different reverse engineering techniques are used to construct GRN. Sometimes, finding suitable cluster accelerate the progress. Basically, the main task in GRN clustering is to find a sub-network which performs a specific task. It is also possible that overlapping may exist in network i.e., same \emph{transcirption factor, regulator} or \emph{gene} may be component of two different sub-networks. In addition, such networks clustering are also used to re-construct GRN. For example, in \cite{godsey}, authors use Bayesian Clustering technique to infer GRN. In GRN, most of the \emph{in silico} benchmark datasets are found in \emph{DREAM} challange \cite{dc}. To apply NEDindex to evaluate such clustering in GRN, we use a data set of DREAM4 which is found in \cite{dream}. In the tested dataset, there are 7115 nodes and 103689 edges. We report the results in Figure \ref{compl}. In this experiment, we use Algorithm \ref{algo1} and vary the size of clusters from 50 to 150. We see in Figure \ref{compl} that both of the metrics jump when cluster size is close to 105. This is because of the strength in clusters i.e., it shows an improved and strongly connected clustering with all cluster components.

\balance
\section{Conclusion}
\label{conclusion}
In this paper, we present NEDindex metric for graph clustering which evaluates the effectiveness of a cluster according to the respective graph. NEDindex has higher value for strongly connected clusters and lower value for weekly connected clusters. We show that our proposed metric can evaluate the effectiveness well and show greater consistency with all kinds of graph clustering \footnote{We show experiments for undirected graphs. We also achieve similar performance for directed graphs.}. We also show some comparative results with other popular graph clustering metrics. Finally, we present two successful experiments of NEDindex for real-world applications (community detection in social networks and sub-networks in GRN). In both of the cases, we see that our proposed NEDindex shows consistency with the variations of cluster size. We would like to use NEDindex, as a future development task, for finding exact and strongly connected communities in social networks.

\end{document}